\newlength\smallfigwidth
\newlength\figwidth
\documentclass[aps,twocolumn,floatfix,amsmath,amssymb,showkeys,showpacs,letterpaper]{revtex4}
\usepackage{amsmath}

\newcommand{\be}{\begin{equation}}
\newcommand{\ee}{\end{equation}}
\newcommand{\bn}{\begin{eqnarray}}
\newcommand{\en}{\end{eqnarray}}
\newcommand{\ii}{{\rm i}}

\usepackage{graphicx}
\usepackage{hyperref}
\usepackage{bm}

\begin{document}

\title{Transverse magnetic field effects on metastable states of magnetic island chains}

  \author{G.\ M.\  Wysin}
  \email{wysin@k-state.edu}
  \homepage{http://www.phys.ksu.edu/personal/wysin}
  \affiliation{Department of Physics, Kansas State University, Manhattan, KS 66506-2601}

\date{October 3, 2024}
\begin{abstract}
{A one-dimensional chain of elongated anisotropic magnetic islands on a nonmagnetic substrate 
with dipolar interactions and an applied magnetic field transverse to the chain is considered.  
With the long axes of the islands perpendicular to the chain, the system allows for
three uniform metastable states: 
(1) tilted dipoles with magnetization at an oblique angle to the chain, 
(2) transverse dipoles with magnetization perpendicular to the chain, and
(3) alternating transverse dipoles with no net magnetization. 
The uniform magnetic field controls their stabilities and
is analyzed for its ability to cause transitions among the states.
The energy and frequency eigenvalues are determined for small-amplitude traveling wave 
deviations of the dipoles. 
The results are summarized in a phase diagram in the field/anisotropy plane, that
highlights the multistable properties of this type of system.
}
\end{abstract}
\maketitle

\section{One-dimensional chains of thin magnetic islands}
\label{intro}
Arrays of elongated magnetic nanoislands fabricated on nonmagnetic substrates, such as two-dimensional (2D) artificial 
spin ices (ASI) \cite{Nisoli13,skjaervo19} and one-dimensional (1D) dipolar chains \cite{Ostman18,Cisternas21} have many 
interesting features, depending on the geometry and competition between shape anisotropy and dipolar interactions.  
In a leading approximation for high aspect ratio islands, their dipole moments are usually approximated as Ising-like
\cite{Ising25}, as shape anisotropy makes them preferentially point close to the long axes of the islands 
(Ref.\ \cite{Wysin15}, Ch.\ 3).
2D artificial spin ices exhibit geometrical frustration \cite{Wang06}, wherein even the lowest energy states cannot 
find a configuration that simultaneously minimizes all of the interaction energies.
%
%
For square lattice ASI, the ground state possesses antiferromagnetic order, as nearest neighbor dipoles alternate 
in direction while trying (but failing) to simultaneously minimize the anisotropy energy and dipolar energy.

A strong applied field that is slowly turned off can leave 2D ASI in a higher energy metastable remanent state, 
that possesses nonzero magnetization, while being locally stable against small perturbations that do not cause a
direct transition back to the ground state.
There has been substantial interest in finding the differences in the magnetic oscillation modes in ground 
and excited states of ASI \cite{Gliga+13,Iacocca+16,Jung+16}, mostly through numerical simulations.
A simplified model was solved analytically for the modes in square lattice ASI in the ground state \cite{Lasnier+20} 
and for remanent states \cite{Wysin23}, assuming Heisenberg-like dipoles with three spin 
components \cite{Heisenberg28,Jiles91}, all without the effects of an applied magnetic field.
The motivation is that different dipolar configurations should be characterized by their dynamic 
modes \cite{Arroo+19,Arora+Das21}.
Further, physical modification of the system by pressure or stress might be useful for modifying its dynamic
properties \cite{Edberg21}.
The mode spectrum also implies the conditions needed for instability of a chosen dipolar configuration, and
this information can be used to predict transitions of the system from an unstable state to a stable state.

\begin{figure}
\includegraphics[width=\figwidth,angle=0]{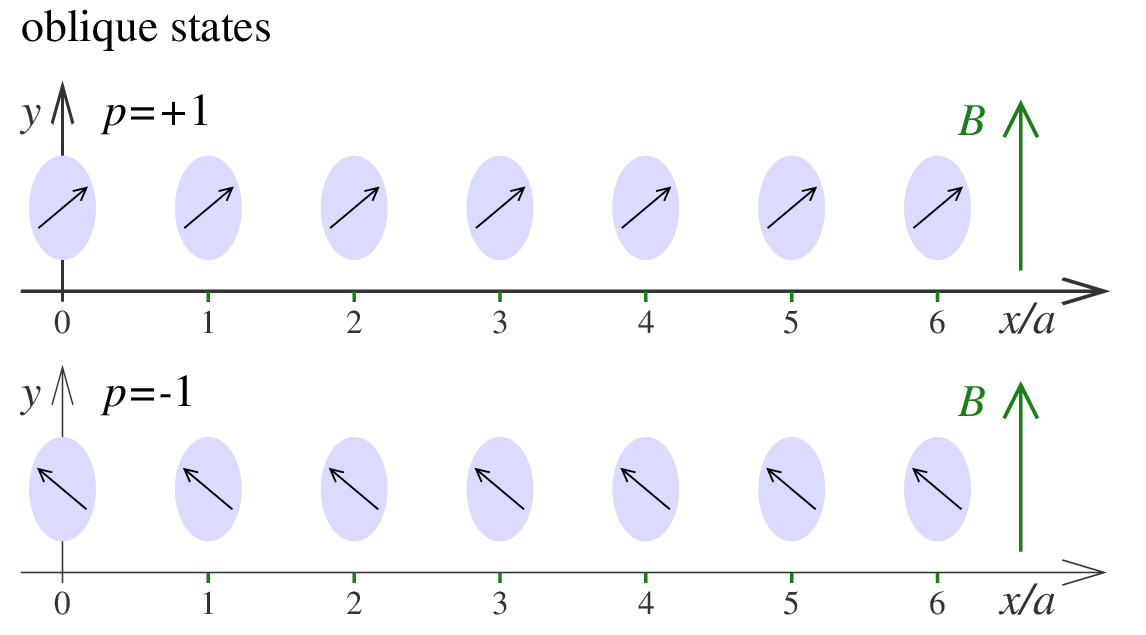}
\caption{\label{xpar-islands} The two possible oblique states, distinguished by dipoles with positive $x$-components 
($p=+1$) or negative $x$-components ($p=-1$). With increasing applied field $B$, the dipolar energy per site changes from 
$-2\zeta_R D$ at $B=0$ (dipoles in $\pm x$ directions) to $+\zeta_R D$ at $\mu B_{\rm max}=2(3\zeta_R D-K_1)$ 
(dipoles parallel to $B$).}
\end{figure}

A 1D chain of islands whose long axes are perpendicular to the chain has been studied \cite{Wysin22} for
some of its similarities to 2D ASI, and this article concerns its properties when a magnetic field {\bf B} is 
applied perpendicular to the chain.  
The system is depicted in Fig.\ \ref{xpar-islands}, where $x$ is along the chain, $y$ is transverse,
and $z$ is perpendicular to the substrate.
It is assumed that the islands are thin perpendicular to the substrate, and only weakly elongated, leading 
to easy-plane anisotropy in the substrate combined with moderate easy-axis anisotropy along the 
longer axes \cite{Wysin+12}.
Assuming single-domain magnetic structure in each nano-sized island, their net dipole moments are Heisenberg-like, 
and their dynamics can be analyzed using Hamiltonian spin dynamics \cite{Wysin+13} (no Ising approximation).
For ${\bf B}=0$, the model allows for two states with remanent magnetization, either parallel to the
chain (called $x$-parallel) or perpendicular to the chain (called $y$-parallel), and one with the 
dipoles alternately pointing perpendicular to the chain (called $y$-alternating). 
The $x$-parallel and $y$-parallel states resemble remanent states of ASI, and the $y$-alternating state
is reminiscent of the alternating ground state of square ASI.
%

The small-amplitude magnetic oscillations were determined for each type of state in Ref.\ \cite{Wysin22}, 
{\em without} a magnetic field, as functions of the anisotropy and dipolar strengths.
The anisotropy relative to dipolar interaction needed to stabilize each type of state was determined.
It was found that $x$-parallel and $y$-alternating states destabilize one into the other with a 
fluctuation at wavevector $q=\pi/a$, where $a$ is the island spacing.
In contrast, the metastable $y$-parallel state destabilizes into $y$-alternating with a fluctuation at 
wavevector $q=0$ (but not \textit{vice-versa}). 
Surprisingly, that transition takes place at an easy-axis anisotropy value where $y$-parallel has the 
same energy as $x$-parallel, even though $x$-parallel is absolutely unstable there, see Fig.\ 9 of
Ref.\ \cite{Wysin22}.

A uniform magnetic field {\bf B} applied transverse to the chain direction has significant consequences.
%
%
To start with, the dipoles in an $x$-parallel state are tilted in the field direction away from 
the chain direction, as in Fig.\ \ref{xpar-islands}, hence they are renamed as \textit{oblique} states now.
Especially, this study shows how an applied field can switch the system among the three states mentioned.
Additionally, the results indicate how it is dynamic fluctuations that determine metastability, which 
is not directly connected to energy differences among the states, as might have been na\"ively assumed.

After finding the new static states when the field is applied, the small-amplitude oscillations
about each state will be determined using undamped Heisenberg spin dynamics.
Based on those spectra, the anisotropy and field strengths at which each state is locally stable 
(\textit{i.e.}, metastable) or unstable will be determined.
The results for all three states taken together will be used to describe possibilities for transitions 
among them.

General features of the states found here could be measurable with magnetic force microscopy \cite{Yue+11} 
or magneto-optic techniques \cite{Haider17,Kimel+22} in similar 1D systems,  such as chains of biomineralized 
magnetosomes \cite{Wittborn+99}, patterned Permalloy elements \cite{Garcia+02a}, Fe nanoparticles \cite{Tang+15}, 
Co$_2$C nanoparticles \cite{Zhang+18}, and nanowire elements \cite{Garcia+02b}.
The results will imply particular jumps or other features in the magnetization plotted versus applied field, 
although the emphasis of this study is on the states' stability as a function of anisotropy and field.

%

\section{The Heisenberg-like macro-dipole model}
For single-domain magnetic islands (small enough with strong internal ferromagnetic exchange), 
the state of one island can be approximated as a single magnetic dipole of fixed magnitude $\mu$,
and arbitrary direction \cite{Wysin+13}.
A particular dipole is denoted $\mu {\bf S}_n$, where ${\bf S}_n$ is a unit spin vector.
It will be convenient to write these spin vectors using planar spherical angles ($\phi_n,\theta_n)$,
where $\phi_n$ is an azimuthal angle in the $xy$-plane and $\theta_n$ is the tilting of a
spin out of the $xy$-plane, i.e.,
\begin{align}
\label{coords}
{\bf S}_n & =(S_n^x,S_n^y,S_n^z) \nonumber \\
& =(\cos\theta_n \cos\phi_n, \cos\theta_n \sin\phi_n, \sin\theta_n).
\end{align}
For dynamics, $S_n^z$ is the momentum conjugate to $\phi_n$.

The islands are assumed to have moderately strong shape anisotropies that tend to
cause their net dipole moment to point within the plane of the island ($xy$) and
prefer to point along its long axis ($y$-direction). 
These preferences are represented mathematically through an easy-plane anisotropy $K_3>0$
and an easy-axis anisotropy $K_1>0$, see \cite{Wysin+12}.  
The $n^{\rm th}$ island's anisotropy contribution to the Hamiltonian is expressed alternatively in 
Cartesian or spherical coordinates as
\begin{align}
H_n^K & \equiv -K_1 \left(S_n^y\right)^2 +K_3 \left(S_n^z\right)^2 
\nonumber \\
& = -K_1\cos^2\theta_n \sin^2\phi_n +K_3\sin^2\theta_n.
\end{align}
Each island interacts with the applied field transverse to the chain, whose contribution to
the Hamiltonian is
\be
H_n^B=-\mu B S_n^y = -\mu B \cos\theta_n\sin\phi_n.
\ee
The chain itself is along the $x$ direction, with nearest-neighbor (NN) pair separation
at lattice constant $a$.
Pairs of dipoles interact via long-range dipolar interactions. 
With $\mu_0$ being the permeability of vacuum, define the energy constant for dipolar interaction of NN-pairs, 
\be
D = \frac{\mu_0 \mu^2}{4\pi a^3}.
\ee
$D$ will be used as the fundamental energy scale in this work.
The dipole interaction is reduced relative to this by the cube of the
separation distance $r$ measured in lattice constants, which is an integer, $k=r/a$. 
With unit vector $\hat{x}$ along the chain direction, the dipolar pair interaction between island 
$n$ and island $n+k$ is 
\begin{align}
H_{n,k}^D & \equiv 
\frac{D}{k^3}
\left[ {\bf S}_n\cdot {\bf S}_{n+k} -3({\bf S}_n\cdot \hat{x})({\bf S}_{n+k}\cdot \hat{x})\right] \nonumber \\
& =  \frac{D}{k^3} \big[\sin\theta_n\sin\theta_{n+k}
+\cos\theta_n \cos\theta_{n+k} \nonumber \\ 
& \times (-2\cos\phi_n\cos\phi_{n+k}+\sin\phi_n\sin\phi_{n+k})\big].  
\end{align}
Then the Hamiltonian for a chain of $N$ dipoles exposed to a uniform magnetic field of
strength $B$ along the $y$-direction is taken as
\be
H = \sum_{n=1}^N \left(H_n^K + H_n^B + \sum_{k=1}^R H_{n,k}^D \right).
\ee
%
%
An upper limit $R$ is used on the range of the dipolar interactions.
When $R=1$, it reverts to a NN-model. 
By summing over all $n$, and only positive values of separation $k$, each dipole-dipole pair
interaction is included once. 
To avoid end effects, it will generally be assumed that $N\rightarrow \infty$ and per-site
energies will be most relevant.
The model calculated with infinite range dipole interactions ($R\to \infty$) will be referred to as 
the long-range dipole (LRD) model.
In a 1D model, convergence of the needed sums is fairly rapid.
Results for the LRD model are different from that for NN interactions only by a slight rescaling of energy 
and frequency eigenvalues, as well as a slight rescaling of the stable state phase diagram obtained later.

\section{Finding stationary states with applied field present}
Initially, uniform stationary states where the dipoles are all parallel are considered.  
The field will tend to tilt the dipoles away from the chain direction.  Therefore we
take all of the spin's angles to be the same unknown values, $(\phi,\theta)$.
The Hamiltonian per site $u=H/N$ becomes
\begin{align}
u = & \sum_{k=1}^{R} \frac{D}{k^3} \left( 1-3 \cos^2 \theta \cos^2\phi \right) 
\nonumber \\
& -K_1 \cos^2 \theta \sin^2 \phi +K_3 \sin^2 \theta -\mu B \cos\theta \sin\phi .
\end{align}
A possible state should minimize this energy.  The derivatives with respect to $\theta$
and $\phi$ must be zero:
\begin{align}
\label{dh1}
\frac{\partial u}{\partial \theta} = & \big[6\zeta_R D \cos^2 \phi+2K_1 \sin^2\phi \nonumber \\
& +2K_3 \cos\theta + \mu B \sin\phi \big] \sin\theta =0,  \\
\label{dh2}
\frac{\partial u}{\partial \phi} = & \big[(6\zeta_R D-2K_1) \cos\theta\sin\phi 
-\mu B \big] \cos\theta\cos\phi =0. 
\end{align}
The dipole sum over range $R\ge 1$ is defined as
\be
\zeta_R \equiv  \sum_{k=1}^{R} \frac{1}{k^3} .
\ee
For infinite range interactions, this becomes a zeta function, 
$\zeta_{\infty}=\zeta(3)\approx 1.2020569$,
while the value for the NN model is simply $\zeta_1=1$.  

\subsection{Oblique states for $B\ne 0$}
Solving Eqs.\ (\ref{dh1}) and (\ref{dh2}), $(0)$ superscripts are used to indicate values 
that minimize the per-site energy, $u$.
The first solution from  Eq.\ (\ref{dh1}) has $\theta^{(0)}=0$, but $S_n^y=\sin\phi^{(0)}$
takes on a value from Eq.\ (\ref{dh2}) that increases with the applied field, satisfying
\be
\label{Sny}
1 \ge S_n^y = \sin\phi^{(0)} = \frac{\mu B/2}{3\zeta_R D-K_1} \ge 0.
\ee
For $B\to 0$, the dipoles are parallel to the chain direction ($\phi^{(0)}=0,\pi$), which are
the $x$-parallel states formerly discussed \cite{Wysin22}. 
A nonzero applied field tilts the dipoles uniformly towards the field direction, as in Fig.\ \ref{xpar-islands},
now referred to as \textit{oblique} states. There are two oblique states: one with positive $x$-components of the dipoles
(a \textit{polarization} along $x$ of $p=+1$) and one with negative $x$-components (polarization $p=-1$).  The 
dipoles have the same $y$-components in both oblique states.

With or without an applied field, the magnetization is saturated, and the field works to rotate it away from the 
chain direction.  In the case of zero applied field, the dipolar energy is minimized  while the $K_1$ anisotropy 
energy is maximized.  A nonzero field brings these energies into competition.  The per-site energy for oblique 
states is found to be
\be
\label{uxp}
u_{\text{oblq}} = -2\zeta_R D -\frac{\left(\mu B/2\right)^2}{3\zeta_R D-K_1}.
\ee
The field lowers the energy of both oblique states equally. 

Based on the factor in the denominator of Eq.\ (\ref{Sny}), an oblique  state can only
exist for $K_1<3\zeta_R D$.  In the absence of a field for the NN-model, stability has been 
shown to require $K_1<D$.  Although a complete stability analysis is presented below, this indicates 
that the applied field is able to extend the range of stability of $x$-parallel states by tilting
the dipoles, forming an oblique state, and lowering the energy. This suggests that under appropriate 
conditions, an oblique state created while a field is applied could be destroyed or transformed to 
another state by turning off the field.

\begin{figure}
\includegraphics[width=\figwidth,angle=0]{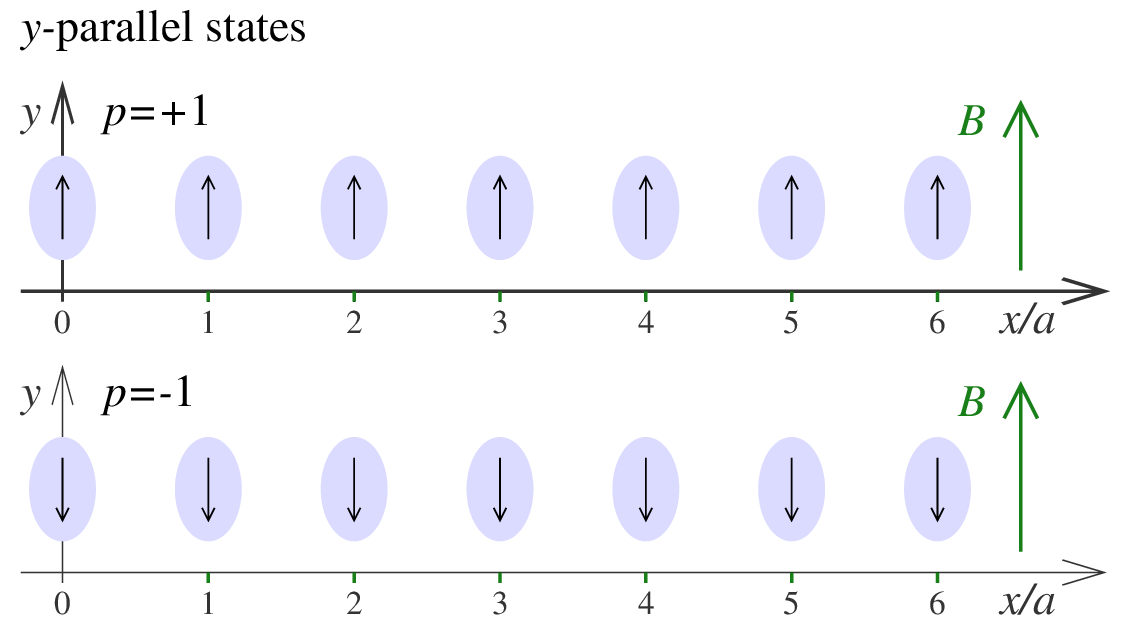}
\caption{\label{ypar-islands} The two possible $y$-parallel states, with dipoles uniformly aligned ($p=+1$) or
anti-aligned ($p=-1$) with the applied field.  The field causes an energy difference between them.}
\end{figure}
%

\subsection{$y$-parallel or transverse states}
A second solution obtained from Eqs.\ (\ref{dh1}) and (\ref{dh2}) has uniform angles
\be
\theta^{(0)} = 0, \quad \phi^{(0)} = p \frac{\pi}{2}, 
\ee
where its polarization along $y$ is $p=\pm 1$. The nonzero spin components are
\be
S_n^y = p = \pm 1.
\ee
The dipoles point perpendicular to the chain direction, either parallel ($p=+1$) or antiparallel 
($p=-1$) to the applied field ${\bf B}$, see Fig.\ \ref{ypar-islands}.  
These are states with a saturated transverse magnetization, called $y$-parallel states in an earlier report 
or $y$-par for short.  The $y$-parallel states minimize the anisotropy energy but not the dipolar energy.  
It is clear that the state polarized parallel to ${\bf B}$
should exhibit greater stability (it minimizes the applied field energy), and conversely, the antiparallel 
state will become unstable once the field surpasses some maximum. 
The per-site energy is found to be
\be
\label{uyp}
u_{y\text{-par}} = \zeta_R D -K_1-p\mu B.
\ee
While the $y$-parallel states take the same perpendicular structure, regardless of the field, their
stability is indeed influenced by the field, which is analyzed below.

\subsection{$y$-alternating states}
A third type of stationary state is possible, primarily due to NN dipolar interactions, where the dipoles
alternate by site, pointing in $\pm y$-directions perpendicular to the chain, called $y$-alternating
or $y$-alt for short.  The structure will produce a dipolar energy that is less than that in 
a $y$-parallel state, but not as low as in oblique states.

\begin{figure}
\includegraphics[width=\figwidth,angle=0]{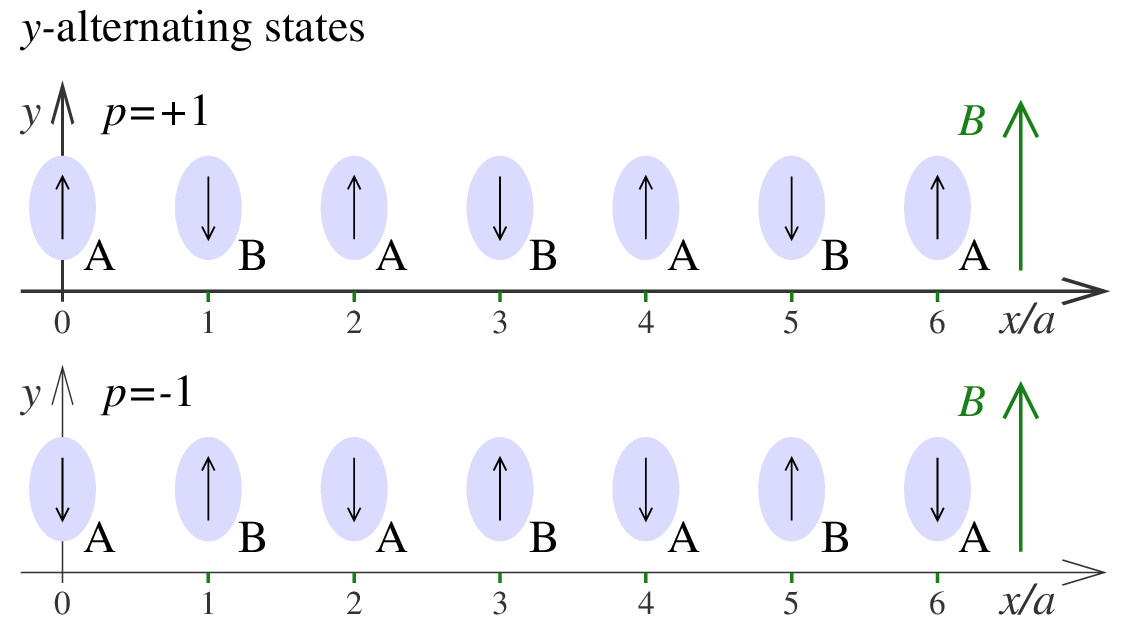}
\caption{\label{yalt-islands} The two possible $y$-alternating states, with dipoles at even sites aligned ($p=+1$) or
anti-aligned ($p=-1$) with the applied field, and dipoles at odd sites in the opposing direction.  
The field does not influence the state energy but does affect the stability.}
\end{figure}

To analyze this configuration, two sublattices A and B are assumed.  Let the A sublattice
be the $n$=even sites and the B sublattice be the $n$=odd sites, and assume spin angles
$(\phi_A,\theta_A)$ and $(\phi_B,\theta_B)$, uniform by sublattice.   Start with nearest-neighbor 
dipole interactions only.  Each dipolar interaction is between A and B sites, at distance $r=a$ or $k=1$. 
\begin{widetext}
\noindent
Averaging over A and B sites, the per-site energy is
\begin{align}
u = & D\big[\cos\theta_A\cos\theta_B\cos(\phi_A-\phi_B)+\sin\theta_A\sin\theta_B
-3\cos\theta_A\cos\theta_B\cos\phi_A\cos\phi_B\big] \nonumber \\
& -\frac{K_1}{2}(\cos^2\theta_A\sin^2\phi_A+\cos^2\theta_B\sin^2\phi_B)
+K_3 (\sin^2\theta_A+\sin^2\theta_B)-\frac{\mu B}{2}(\cos\theta_A\sin\phi_A+\cos\theta_B\sin\phi_B).
\end{align}
\end{widetext}
This is now minimized with respect to the four angles.  A brief calculation shows that
\be
\frac{\partial u}{\partial \theta_A} = \frac{\partial u}{\partial \theta_B}=0 
\ee
is satisfied by 
\be
\theta_A^{(0)} = \theta_B^{(0)}= 0.
\ee
Statically, the spins remain in the $xy$-plane.  The remaining equations are of this form:
\begin{align}
\frac{\partial u}{\partial\phi_A} = & D\big[-\sin(\phi_A-\phi_B)+3\sin\phi_A\cos\phi_B\big]
\nonumber \\
& -K_1 \sin\phi_A\cos\phi_A-\frac{\mu B}{2}\cos\phi_A = 0.
\end{align}
Letting $\phi_B=\phi_A$ recovers the oblique and $y$-parallel states.  Instead, trying
$\phi_B=-\phi_A$ produces
\be
\left[(D-K_1)\sin\phi_A-\tfrac{1}{2}\mu B\right] \cos\phi_A =0,
\ee
and this is solved by 
\be
\phi_A^{(0)} = p\frac{\pi}{2} = -\phi_B^{(0)}, \quad p=\pm 1.
\ee
There are two degenerate solutions corresponding to the two choices of $p$. 
The alternating angles correspond to dipoles alternating in direction by site,
\be
S_n^y = (-1)^n p.
\ee
The two possible $y$-alt states are shown in Fig.\ \ref{yalt-islands}.
While they exhibit antiferromagnetic order, the interaction is dipolar and the structure has 
nothing to do with antiferromagnetism.  Hence the preferred name is $y$-alternating or just $y$-alt. 

One can assume the same alternating structure might exist even with
longer range dipole interactions, and later investigate its stability.  Consider the
energy per site.  The dipolar contributions around a central A-site alternate in sign, 
as they change between AB bonds ($\phi_A-\phi_B=p\pi$) and AA bonds ($\phi_A-\phi_A=0$).  
Thus, the energy per site for $y$-alternating states is seen to be
\be
\label{uyalt}
u_{y\text{-alt}} = D\sum_{k=1}^{R} \frac{(-1)^k}{k^3} -K_1.
\ee
The sum needed here can be expressed by separating even and odd contributions, which involve
\begin{align}
s_{\rm e}=\sum_{k=2,4,6...}^{R} \frac{1}{k^3} & = \tfrac{1}{8} \sum_{k=1}^{R} \frac{1}{k^3} = \tfrac{1}{8}\zeta_R,
\nonumber \\
s_{\rm o}=\sum_{k=1,3,5...}^{R} \frac{1}{k^3} & = \sum_{k=1}^{R} \frac{1}{k^3}-\sum_{k=2,4,6...}^{R} \frac{1}{k^3} 
= \tfrac{7}{8} \zeta_R.
\end{align}
Then the sum needed is 
\be
\sum_{k=1}^{R} \frac{(-1)^k}{k^3} = -\tfrac{7}{8}\zeta_R+\tfrac{1}{8}\zeta_R = -\tfrac{3}{4}\zeta_R,
\ee
and the $y$-alt state energy per site is independent of $p$ and the applied field,
\be
\label{uya}
u_{y\text{-alt}} = -\tfrac{3}{4} \zeta_R D -K_1.
\ee
Even though $B$ is absent in this expression, the field strength determines the stability of $y$-alt states.

\section{Dynamics and stability}
Suppose there is a gyromagnetic ratio $\gamma_{\rm e}$ that converts angular momenta ${\bf L}_n$ into magnetic 
dipole moments via $\vec\mu_n = \gamma_{\rm e}{\bf L}_n$.  Then the undamped free dynamics of a magnetic dipole system 
follows a torque equation for the time derivative of ${\bf L}_n$ (see Ref.\ \cite{Wysin15}, Ch.\ 5), 
\be
\frac{d{\bf L}_n}{dt} = \vec\tau_n = \vec\mu_n\times {\bf B}_n^{\rm eff} ,
\ee
where the Hamiltonian involves effective fields ${\bf B}_n^{\rm eff}$ at each site, 
$H=-\sum_n \vec\mu_n \cdot {\bf B}_n^{\rm eff}$.  This gives
\be
\frac{1}{\gamma_{\rm e}} \frac{d\vec\mu_n}{dt} 
= \vec\mu_n \times \left(-\frac{\partial H}{\partial \vec\mu_n}\right).
\ee
Transforming the dipoles to the spherical coordinates in (\ref{coords}), the mechanics is that
where $\phi_n$ are generalized coordinates and $\sin\theta_n$ are the corresponding conjugate
momenta.  The dynamics obeys Hamiltonian equations, 
\begin{align}
\label{HE}
\frac{\mu}{\gamma_{\rm e}} \frac{d}{dt}\phi_n & = \frac{\partial H}{\partial \sin\theta_n},
\nonumber \\
\frac{\mu}{\gamma_{\rm e}} \frac{d}{dt}\sin\theta_n & = -\frac{\partial H}{\partial \phi_n}.
\end{align}


\subsection{Linearization of $H$}
In practice, we consider the dynamics linearized around the three types of states described above,
all of which have $\theta^{(0)}=0$.   At each site of the chain, let the in-plane angle be replaced as
$\phi_n\rightarrow \phi^{(0)}+\phi_n$, where $\phi_n \ll 1$ now represents a small deviation from the equilibrium 
value.  Similarly, with $\theta^{(0)}=0$,  use $\theta_n \ll 1$ to represent a small deviation from zero.  Then the 
Hamiltonian is expanded to quadratic orders in these deviations, and from there the linearized dynamics 
and stability can be determined.

The sets of in-plane and out-of-plane deviations can be represented by row vectors of the $\phi_n$ and $\theta_n$ angles,
\be
\psi_{\phi}^{\dagger}=(\phi_1,\phi_2,\phi_3...), \quad
\psi_{\theta}^{\dagger}=(\theta_1,\theta_2,\theta_3...).
\ee
The Hamiltonian is expanded in terms of these as 
\be
H=H^{(0)}+H^{(1)}+H^{(2)}.
\ee
$H^{(0)}$ is the minimized state energy that does not depend on the deviations $\phi_n, \theta_n$. 
$H^{(1)}$ is the terms linear in $\phi_n$ and $\theta_n$, which is zero because the state is an energy 
minimum.  $H^{(2)}$ is the deviation energy, a double quadratic form in the deviations,
\begin{align}
\label{qf}
H&^{(2)}  = H_{\phi}+H_{\theta}, \nonumber \\
H&_{\phi}  = \psi_{\phi}^{\dagger} \bm{M}_{\phi} \psi_{\phi}, \quad
 H_{\theta} = \psi_{\theta}^{\dagger} \bm{M}_{\theta} \psi_{\theta}.
\end{align}
The elements of the matrices $\bm{M}_{\phi}$ and $\bm{M}_{\theta}$ come from expanding around each of the three states.
These matrices determine the traveling wave fluctuations of the system, and instabilities of those waves 
(such as imaginary eigenfrequencies) signal instability of a state.

\subsection{Linearized dynamics, energy eigenvalues, instabilities}
With the Hamiltonian linearized and described by matrices $\bm{M}_{\phi}$ and $\bm{M}_{\theta}$, even 
including long-range dipole interactions, the dynamic equations of motion (\ref{HE}) take the form,
\begin{align}
\label{LRDmotion}
\frac{\mu}{\gamma_{\rm e}}\dot{\phi}_n & = +2 \sum_{k=-R}^{+R} M_{\theta,n,n+k}\theta_{n+k}, \nonumber \\
\frac{\mu}{\gamma_{\rm e}}\dot{\theta}_n &  = -2\sum_{k=-R}^{+R} M_{\phi,n,n+k}\phi_{n+k},
\end{align}
where dot signifies time derivative. The sums include diagonal or on-site matrix elements ($k=0$) as well as dipole pair
interactions at separations $k \ne 0$ out to range $R$ in both directions.
%
This is equivalent to the pair of matrix equations,
\be
\frac{\mu}{\gamma_{\rm e}} \dot{\psi}_{\phi} = +2\bm{M}_{\theta}\psi_{\theta}, \qquad
\frac{\mu}{\gamma_{\rm e}} \dot{\psi}_{\theta} = -2\bm{M}_{\phi}\psi_{\phi},
\ee
where $\psi_{\phi}$ and $\psi_{\theta}$ are column vectors of the angles.
In a system with a single sublattice, these equations allow for travelling waves
with small amplitudes $a_{\phi}$ and $a_{\theta}$ and frequency $\omega$, 
\be
\label{awave}
\phi_n = a_{\phi} e^{\ii (q n a-\omega t)}, \qquad \theta_n = a_{\theta} e^{\ii (q n a-\omega t)}. 
\ee
where the allowed wavevectors for periodic boundary conditions are
\be
q\equiv \frac{2\pi m}{Na}, \quad m=0,1,2...(N-1).
\ee
The equations of motion (\ref{LRDmotion}) condense into a pair of equations involving only
the two amplitudes, 
\begin{align}
-\ii \omega a_{\phi} &= +2 \frac{\gamma_{\rm e}}{\mu} \lambda_{\theta} a_{\theta}, \nonumber \\
-\ii \omega a_{\theta} &= -2 \frac{\gamma_{\rm e}}{\mu} \lambda_{\phi} a_{\phi}.
\end{align}
These are expressed in terms of energy eigenvalues $\lambda_{\phi}$ and $\lambda_{\theta}$ of 
the $\bm{M}_{\phi}$ and $\bm{M}_{\theta}$ matrices, defined in a usual way,
\be
\label{ll}
\bm{M}_{\phi} \psi_{\phi} = \lambda_{\phi} \psi_{\phi}, 
\quad 
\bm{M}_{\theta} \psi_{\theta} = \lambda_{\theta} \psi_{\theta}, 
\ee
where $\psi_{\phi}$ and $\psi_{\theta}$ are column eigenvectors composed from the site angles. 
Therefore, the frequency for a travelling wave at wavevector $q$ is
\be
\label{wqll}
\omega(q) = 2\frac{\gamma_{\rm e}}{\mu} \sqrt{\lambda_{\phi}\lambda_{\theta}}.
\ee
Results are given for frequencies in terms of a frequency unit based on the NN-dipolar coupling frequency,
\be
\label{del1}
\delta_1 \equiv \frac{\gamma_{\rm e}D}{\mu},
\ee
as in Fig.\ \ref{wxlrd-10} and the other $\omega(q)$ dispersion relation plots.

Assuming inversion symmetry, the eigenvalue problems can be written in a form,
\be
M_{\phi,n,n}\phi_n+\sum_{k=1}^R M_{\phi,n,n+k}(\phi_{n+k}+\phi_{n-k}) = \lambda_{\phi} \phi_n.
\ee 
Then the energy eigenvalues obtained for the assumed travelling waves can
be expressed as 
\begin{align}
\label{lf}
\lambda_{\phi}(q) & 
= M_{\phi,n,n}+2\sum_{k=1}^{R} M_{\phi,n,n+k}\cos kqa, \nonumber \\
\lambda_{\theta}(q) & 
= M_{\theta,n,n}+2\sum_{k=1}^{R} M_{\theta,n,n+k} \cos kqa.
\end{align}
The expressions apply to the translationally invariant oblique and $y$-parallel states.  
For the $y$-alternating states, a similar procedure but with a two-sublattice wave
assumption is applied in Sec.\ \ref{Lin-yalt}.

Instabilities of a given state will be considered due to varying the anisotropy constants or 
the applied field.  An instability is associated with an arbitrary fluctuation that
lowers the energy.  That is indicated when one of the energy eigenvalues $\lambda_{\phi}$ or 
$\lambda_{\theta}$ becomes zero or even negative at some wavevector. 
If either eigenvalue goes to zero or becomes negative, then the frequency $\omega(q)$ 
also goes to zero or becomes imaginary.

This method determines the presence of 
any dynamic instability in the chosen state.  Further, the wavevector where that occurs
gives an indication of the unstable change in structure of the state, and hence is
a guide towards what structure will result if the instability takes over the dynamics.
These properties are determined separately for oblique, $y$-parallel and $y$-alternating
states, after determining their dynamic matrices $\bm{M}_{\phi}$ and $\bm{M}_{\theta}$.

\subsection{Linearized analysis of oblique states}
For oblique states, small deviations of the dipoles away from equilibrium are considered.
The analysis is complicated by the fact that in equilibrium the dipoles are
tilted away from the $x$-axis when a field is applied along the $y$-axis.
Energy changes are considered when small deviations take place relative to that oblique direction.

\subsubsection{Expanding $H$ for oblique states}
For an oblique state, the in-plane angles $\phi_n$ are replaced by $\phi_n \rightarrow \phi^{(0)}+\phi_n$,
where $\phi^{(0)}$ is the equilibrium value given in Eq.\ (\ref{Sny}) and $\phi_n$ is now the {\em deviation} from that. 
The out-of-plane deviations are $\theta_n$. To facilitate the algebra, use a notation,
\be
s_0=\sin\phi^{(0)}, \quad c_0 = \cos\phi^{(0)}.
\ee
Then the expansion of a dipole pair interaction in $H$ for site $n$ interacting with site $n+k$,  
to quadratic order in deviations, is
\begin{align}
H_{n,k}^D &
\approx \frac{D}{k^3} \Big[ (c_0^2-\tfrac{1}{2}s_0^2)(-2+\phi_n^2+\phi_{n+k}^2+\theta_n^2+\theta_{n+k}^2)
\nonumber \\
& +(c_0^2-2s_0^2)\phi_n \phi_{n+k} +\theta_n\theta_{n+k} +3c_0 s_0 (\phi_n+\phi_{n+k}) \Big].
\end{align}
This is summed over all $n$ but only with $k\ge 1$ to count all pairs. 
The anisotropy energy at a site is
\begin{align}
H_n^K & \approx -K_1[s_0^2(1-\phi_n^2-\theta_n^2)+c_0^2 \phi_n^2+2c_0 s_0 \phi_n] +K_3\theta_n^2.
\end{align}
The field term is
\begin{align}
H_n^B \approx -\mu B \left[s_0(1-\tfrac{1}{2}\phi_n^2-\tfrac{1}{2}\theta_n^2)+c_0 \phi_n\right].
\end{align}
The combination of all these parts gives the per-site Hamiltonian, and in this case we see zeroth, first,
and quadratic order parts, $H_n \approx H_n^{(0)}+H_n^{(1)}+H_n^{(2)}$. The zeroth is
\be
H_n^{(0)} = \left(\sum_{k=1}^{R} \frac{1}{k^3}\right) D(-2c_0^2+s_0^2) -K_1 s_0^2 -\mu B s_0.
\ee
Minimized with respect to $s_0$, the equilibrium results of Eq.\ (\ref{Sny}) for $s_0$ and $\phi^{(0)}$ are 
recovered.  In first order, there is
\be
H_n^{(1)} = \left[ \left(\sum_{k=1}^{R} \frac{1}{k^3}\right) 6D s_0 -2K_1 s_0 -\mu B \right] c_0 \phi_n.
\ee
This is identically zero when the equilibrium value of $s_0$ is inserted, as it should be
in a minimizing state. Finally there is the quadratic part,
\begin{align}
H_n^{(2)} & = D \sum_{k=1}^{R} \frac{1}{k^3} \big[ (2-3s_0^2)(\phi_n^2+\theta_n^2)
+(1-3s_0^2)\phi_n\phi_{n+k} \nonumber \\ & +\theta_n\theta_{n+k}\big]
-K_1 [(c_0^2-s_0^2)\phi_n^2-s_0^2\theta_n^2]+K_3\theta_n^2 \nonumber \\
& +\tfrac{1}{2}\mu B s_0 (\phi_n^2+\theta_n^2).
\end{align}
The $\phi$ and $\theta$ contributions are completely decoupled, allowing us to write 
$H^{(2)}=H_{\phi}+H_{\theta}$ and expressing these in the matrix notation of Eq.\ (\ref{qf}).
The elements of matrices $\bm{M}_{\phi}$ and $\bm{M}_{\theta}$ can be determined, assuming
dipole interactions out to maximum range $R$.  Being the coefficients of $\phi_n^2$ and $\theta_n^2$ in 
$H_n^{(2)}$, the on-site elements are surprisingly simple when using the equilibrium value $s_0$,
\be
M_{\phi,n,n}  
= 2\zeta_R D-K_1 c_0^2, \quad
M_{\theta,n,n}  
 = 2\zeta_R D+K_3.
\ee
The inter-site elements are also simple, being half the coefficients of pairs of angles in $H_n^{(2)}$,
\be
M_{\phi,n,n+k} = \frac{D}{2k^3}(1-3s_0^2), \quad M_{\theta,n,n+k} = \frac{D}{2k^3}.
\ee
Note that the applied field only affects the in-plane parts.

\subsubsection{Energy and frequency eigenvalues of oblique states}
\begin{figure}
\includegraphics[width=\figwidth,angle=0]{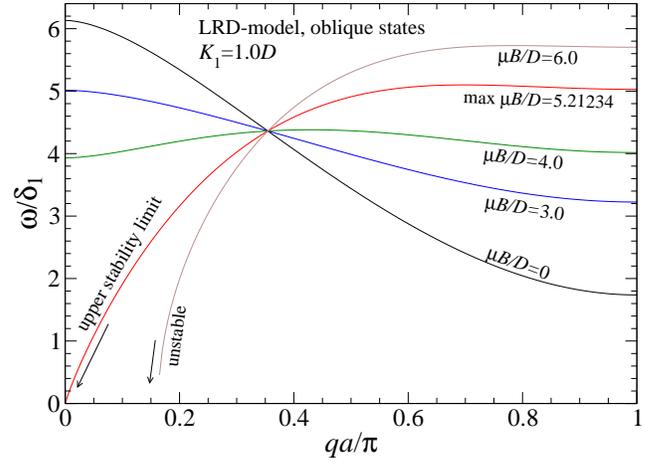}
\caption{\label{wxlrd-10} Mode frequencies from Eq.\ (\ref{wqll}) in the LRD-model for oblique states with
$K_1=1.0 D$, $K_3=0$, with a transverse applied field strength as indicated. The frequency unit $\delta_1$ is
defined in Eq.\ (\ref{del1}). At $\mu B_{\rm max} \approx 5.21234 D$,
the $q=0$ frequency becomes zero and that is the limit for stability.  For $\mu B>5.21234 D$, imaginary values of
$\omega$ are present near low $q$ and the state is absolutely unstable.}
\end{figure}
The energy eigenvalues associated with small wave-like in-plane deviations ($\lambda_{\phi}$) and for small
wave-like out-of-plane deviations ($\lambda_{\theta}$) control the basic dynamics, and as well,
determine the states' stability. 
Using the assumed wave deviation in (\ref{awave}), the eigenvalues of matrices $\bm{M}_{\phi}$
and also $\bm{M}_{\theta}$ are given by the expression in Eq.\ (\ref{lf}).  Using the matrix
elements just found, one has
\begin{align}
\label{oblams}
\lambda_{\phi}(q) & 
= 2\zeta_R D-K_1 c_0^2+D(1-3s_0^2)\sum_{k=1}^{R} \frac{1}{k^3} \cos kqa, \nonumber \\
\lambda_{\theta}(q) & 
= 2\zeta_R D+K_3+D\sum_{k=1}^{R} \frac{1}{k^3} \cos kqa.
\end{align}
The sum in each expression is a finite-range Clausen function \cite{Clausen} of order 3, 
\be
{\rm Cl}_{R,3}(qa)=  \sum_{k=1}^R \frac{\cos kqa}{k^3}.
\ee
In the limit $R\to \infty$, notable values are 
\begin{align}
{\rm Cl}_{3}(0) &= \zeta(3) \approx 1.2020569, \nonumber \\ 
{\rm Cl}_{3}(\pi) &= -\tfrac{3}{4}\zeta(3) \approx -0.901542677\ .
\end{align}
Only $\lambda_{\phi}(q)$ is affected by the applied field.

Frequencies $\omega(q)$ as obtained from (\ref{wqll}) are shown in Fig.\ \ref{wxlrd-10} for $K_1=1.0 D$, $K_3=0$,
and a range of applied field values $\mu B/D$ from 0 to 6.  This is an anisotropy value that does not require an
applied field being present for stability.  There is a $q$-value where all the dispersion relations cross, 
regardless of the field value.  When $\mu B$ reaches the maximum allowed value, instability takes place at 
$q=0$.  That destabilizes the system to transform into a $y$-parallel state, which is the only available 
structure connected by a $q=0$ perturbation.

\begin{figure}
\includegraphics[width=\figwidth,angle=0]{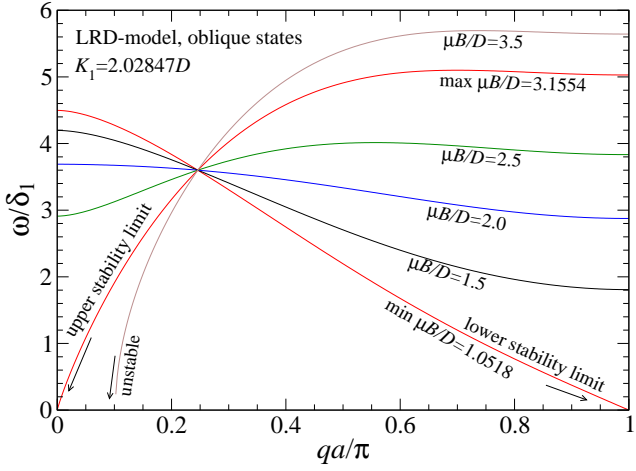}
\caption{\label{wxlrd-20} Mode frequencies from Eq.\ (\ref{wqll}) in the LRD-model for oblique states with
$K_1=2.02847 D$, $K_3=0$, with a transverse applied field strength as indicated.  The field  $\mu B/D$ must be
between required minimum and maximum values for stability at this anisotropy strength. The
state transforms to $y$-par if $\mu B> \mu B_{\rm max}$ and to $y$-alt if $\mu B< \mu B_{\rm min}$. This
anisotropy value is that for the triple point of the system, see Fig.\ \ref{phases_lrd}.}
\end{figure}

Further results are shown in Fig.\ \ref{wxlrd-20} for $K_1=2.02847D$, $K_3=0$, where a field {\em is needed} to stabilize the
state.  One sees a limited range of field that can accomplish that.  At this value of $K_1$, the maximum possible
field where the oblique state is maintained stable is $\mu B_{\rm max} \approx 3.1554 D$.  This pair of $(K_1,B)$ values
constitutes a type of triple point, where oblique, $y$-parallel, and $y$-alternating states are  all nominally
stable, see Fig.\ \ref{phases_lrd} later. At the maximum allowed field, the instability takes place at $q=0$,
showing the tendency to transform into a $y$-parallel state.  To the contrary, at the minimum required field,
the instability takes place at $qa=\pi$, indicating a transformation into a $y$-alt state.

\subsubsection{General stability of oblique states}
Stability of oblique states has two requirements.  The first is that $\lambda_{\phi}(q)>0$
for any value of wavevector $q$.  Physically, the tendency of an oblique state to destabilize
occurs at $qa\approx \pi$, because that type of perturbation deforms it towards an available $y$-alternating 
state.  This can also be seen by realizing that $\lambda_{\phi}(q)$ becomes smallest at $qa=\pi$, because that
makes the terms with odd $k$ in the Clausen sums in Eq.\ \ref{oblams} negative. Enforcing this first constraint, 
stability requires
\be
\sin^2\phi^{(0)} > \frac{K_1-[2\zeta_R+{\rm Cl}_{R,3}(\pi)]D}{K_1-3 {\rm Cl}_{R,3}(\pi) D}.
\ee
But the equilibrium angle is determined by the applied field, so using (\ref{Sny}) this translates into a requirement
on the applied field, 
\be
\label{xparb1}
\mu B > 2(3\zeta_R D-K_1) \sqrt{\frac{K_1-[2\zeta_R+{\rm Cl}_{R,3}(\pi)]D}{K_1-3{\rm Cl}_{R,3}(\pi)D}}.
\ee
For the LRD model, this expression requires  $K_1 > 2\zeta(3)+{\rm Cl}_3(\pi) = (5/4)\zeta D \approx 1.50257\ D$, 
which is the zero-field anisotropy limit, see the $x$-parallel data in Fig.\ 9 of Ref.\ \cite{Wysin22}.

The second requirement for stability is that $s_0 <1$, otherwise, the system would transform
into $y$-parallel.  This would take place at $qa=0$, or, applying the result (\ref{Sny}) and
solving for the allowed field, 
\be
\label{xparb2}
\mu B < 2(3\zeta_R D-K_1).
\ee
The right hand side is the upper limit for allowed applied field. It also implies
that oblique states do not exist if $K_1>3\zeta_R D$, for any applied field.  
\begin{figure}
\includegraphics[width=\figwidth,angle=0]{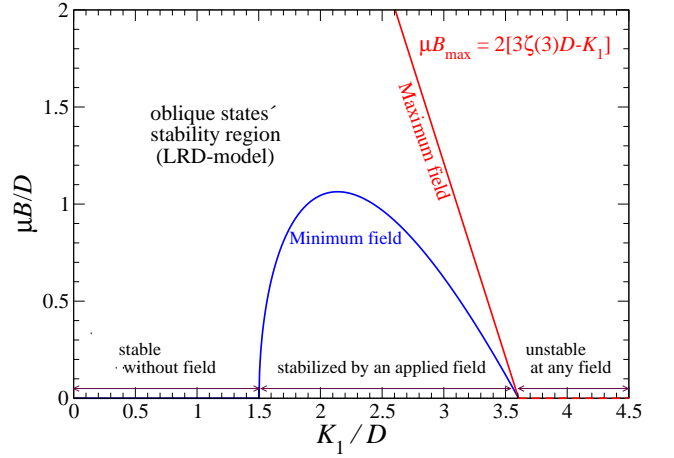}
\caption{\label{muB_lrd_xpar} For oblique states with LRD interactions, the minimum 
and maximum applied field strength $\mu B$ required for stabilization, as a function 
of the in-plane anisotropy $K_1$, both relative to dipolar coupling $D$, as given in Eqs.\ 
(\ref{xparb1}) and (\ref{xparb2}) with $R\to\infty$.}
\end{figure}
In the LRD model, the region with $1.50257 D < K_1 < 3.60617 D$ requires a field for stabilization.
The limited range of allowed field and anisotropy for the LRD model is indicated in Fig. \ref{muB_lrd_xpar}.

These results show two things: (1) An applied field extends the range of stability of oblique 
states to higher in-plane anisotropy values ($K_1$) compared to $x$-parallel states with no applied field, and 
(2) there is a limited range of applied field that will stabilize oblique states, which depends only on the 
in-plane anisotropy. Outside of this required range of $\mu B$, the system will either transform to a 
$y$-alternating state ($\mu B$ too small) or to a $y$-parallel state ($\mu B$ too large).

\subsection{Linearized analysis of $y$-parallel states}
Continue with $y$-parallel states, and consider their energy changes when small deviations of
the dipoles take place, followed by an analysis of stability and dynamics.

\subsubsection{Expanding $H$ for $y$-parallel states}
For a $y$-parallel state with replacement $\phi_n \rightarrow p\frac{\pi}{2}+\phi_n$, the Hamiltonian can 
be expanded to quadratic order in the deviations $\phi_n,\theta_n$, as follows.  The 
dipolar pair energy of site $n$ interacting with site $n+k$ is 
\begin{align}
H_{n,k}^D  = 
\frac{D}{k^3} \Big( 1 & -\tfrac{1}{2}\phi_n^2-\tfrac{1}{2}\phi_{n+k}^2-2\phi_n \phi_{n+k} 
\nonumber \\ 
& -\tfrac{1}{2}\theta_n^2-\tfrac{1}{2}\theta_{n+k}^2+\theta_n\theta_{n+k} \Big).
\end{align}
This is summed over all $n$ and only $k\ge 1$ to get the total dipolar energy.
In addition, there are anisotropy terms,
\be
H_n^K=-K_1\left(1-\phi_n^2-\theta_n^2\right)+K_3\theta_n^2,
\ee 
and field terms,
\be
H_n^B = -p\mu B\left(1-\tfrac{1}{2}\phi_n^2-\tfrac{1}{2}\theta_n^2\right).
\ee
The combination of these parts gives the per-site Hamiltonian with zeroth and quadratic order parts,  
$H_n=H_n^{(0)}+H_n^{(2)}$,
\be
H_n^{(0)} = D\sum_{k=1}^R \frac{1}{k^3}-K_1-p\mu B =\zeta_R D-K_1-p\mu B,
\ee
which agrees with Eq.\ (\ref{uyp}), and 
\begin{align}
H_n^{(2)} &= D\sum_{k=1}^R \frac{1}{k^3} \left(-\phi_n^2-2\phi_n\phi_{n+k}-\theta_n^2+\theta_n\theta_{n+k}\right)
\nonumber \\
& +K_1 \phi_n^2+(K_1+K_3) \theta_n^2 +\tfrac{1}{2} p\mu B(\phi_n^2+\theta_n^2).
\end{align}
In-plane and out-of-plane deviation energies are completely decoupled.  The on-site matrix elements of 
$\bm{M}_{\phi}$ and $\bm{M}_{\theta}$ are the coefficients of $\phi_n^2$ and $\theta_n^2$ in $H_n^{(2)}$,
\begin{align}
M_{\phi,n,n} &= -\zeta_R D+K_1+\tfrac{1}{2} p\mu B, \nonumber \\
M_{\theta,n,n} &= -\zeta_R D+K_{13}+\tfrac{1}{2} p\mu B, 
\end{align}
where the net out-of-plane anisotropy strength is
\be 
K_{13}\equiv K_1+K_3.
\ee
The inter-site matrix elements are half the coefficients of $\phi_n\phi_{n+k}$ and $\theta_n\theta_{n+k}$
in $H_n^{(2)}$,
\be
M_{\phi,n,n+k}=-\frac{D}{k^3}, \quad M_{\theta,n,n+k} = \frac{D}{2k^3}.
\ee
The system is symmetric along the chain, so there are the same matrix elements for $n,n-k$ bonds.

\subsubsection{Energy and frequency eigenvalues of $y$-parallel states}
The energy eigenvalues for small deviations control dynamics and determine stability.
For an in-plane deviation, the matrix elements substituted into Eq.\ (\ref{lf}) give the eigenvalues,
\be
\label{lfy}
\lambda_{\phi}(q) = -\zeta_R D -2D \sum_{k=1}^R \frac{\cos kqa}{k^3} +K_1+\tfrac{1}{2}p\mu B.
\ee
A similar expression is found for the eigenvalues associated with out-of-plane deviations,
\be
\label{lty}
\lambda_{\theta}(q) = -\zeta_R D +D \sum_{k=1}^R \frac{\cos kqa}{k^3} +K_{13}+\tfrac{1}{2}p\mu B.
\ee
Both increase linearly with the field factor, $p\mu B$. 

\begin{figure}
\includegraphics[width=\figwidth,angle=0]{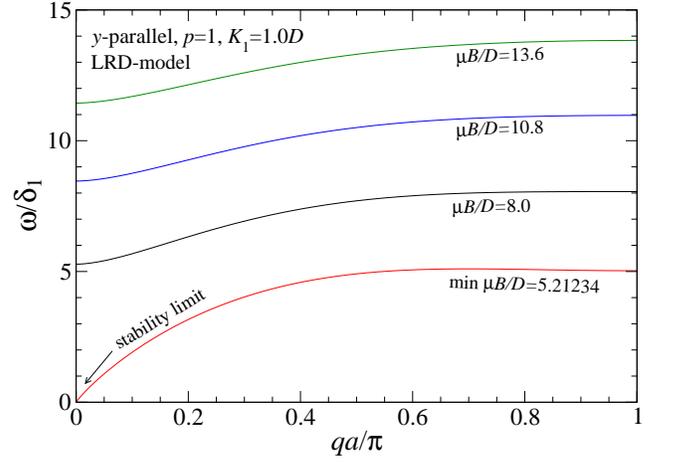}
\caption{\label{wylrd-10} Mode frequencies from Eq.\ (\ref{wqll}) in the LRD-model ($R\to \infty$) for $y$-parallel states 
with $K_1=1.0 D$, $K_3=0, p=+1$, with transverse applied field strength as indicated.  The field  $\mu B/D$ must be
above a required minimum for $p=+1$ $y$-parallel stability, but any large positive field is allowed. If  $\mu B/D$
is below the stability limit, the chain will transform into an oblique state because the $q=0$ fluctuations
will only connect to that state structure at this value of $K_1$.}
\end{figure}

A first example of the mode frequencies resulting from Eq.\ (\ref{wqll}) is shown in Fig.\ \ref{wylrd-10}
for $K_1=1 D$ with a $y$-parallel state with polarization $p=+1$.  A large positive field (aligned
with the polarization direction of the state) will not destabilize it.  To the contrary, there is a
minimum positive applied field (same as the polarization direction, of value $\mu B/D \approx 5.21234$) 
below which the state is destabilized by $q=0$ fluctuations, presumably into an oblique state, that being
the only available stable state for this low value of $K_1$.

A second example of the mode frequencies is shown in Fig.\ \ref{wylrd-50}
for $K_1=5 D$ and a $y$-parallel state with polarization $p=+1$.  A large positive field (aligned
with the polarization direction of the state) will not destabilize it.  To the contrary, there is a
minimum applied field (opposite to the polarization direction, or negative, of value $\mu B/D\approx -2.787$) 
below which the state is destabilized by $q=0$ fluctuations into the other $y$-parallel state with $p=-1$. 
(The only other available state would be $y$-alt, but that does not connect to $y$-parallel via a $q=0$ fluctuation.)
Note the vivid similarity to Fig.\ \ref{wylrd-10}: the curves are nearly the same shapes, because equal
field increments are used in the curves, starting from the minimum required for stability at the
applied value of $K_1$, see Eqs.\ (\ref{lfy}) and (\ref{lty}), where the frequencies shift equivalently with
a change in $K_1$ as with a change in $\frac{1}{2}p\mu B$.

\begin{figure}
\includegraphics[width=\figwidth,angle=0]{wylrd-50}
\caption{\label{wylrd-50} Mode frequencies from Eq.\ (\ref{wqll}) in the LRD-model for $y$-parallel states with
$K_1=5 D$, $K_3=0, p=+1$, with transverse applied field strength as indicated.  The field  $\mu B/D$ must be
above a required minimum for $p=+1$ $y$-parallel stability, but any large positive field is allowed. If  $\mu B/D$
is below the stability limit, the state will transform into $p=-1$ $y$-parallel because the $q=0$ fluctuations
will only connect to that state structure.}
\end{figure}
%

\subsubsection{General stability of $y$-parallel states}
The $y$-parallel states are stable if the energy eigenvalues $\lambda_{\phi}$ and $\lambda_{\theta}$
remain positive, for any wave-like deviation (i.e., for all possible wavevectors $q$). 
At the minimum applied field for which $y$-parallel exists, the frequency goes to zero at $q\approx 0$, see Figs.\ 
\ref{wylrd-10} and \ref{wylrd-50}.  This implies that a $y$-parallel state might have a long-wavelength deviation 
that would tend to rotate it into another allowed state  Noting that Cl$_{R,3}(0)=\zeta_R$, the eigenvalues at $q=0$ are
\begin{align}
\lambda_{\phi}(0) & = K_1+\tfrac{1}{2}p\mu B-3\zeta_R D, \nonumber \\
\lambda_{\theta}(0) & = K_{13}+\tfrac{1}{2}p\mu B.
\end{align}
Assuming these must be positive for stability, they give requirements on the applied field,
\begin{align}
\lambda_{\phi}(0) &>0 \implies p \mu B > 2(3\zeta_R D -K_1), \nonumber \\
\lambda_{\theta}(0) &>0 \implies p\mu B > -2 K_{13}. 
\end{align}
The requirement from $\lambda_{\phi}(0)>0$ is more restrictive and is the deciding factor.  
Therefore, when the dipoles are aligned with {\bf B}, the allowed field range is
\be
\label{yparb1}
\mu B > 2(3\zeta_R D-K_1), \quad p=+1.
\ee 
If the $y$-parallel state has dipoles pointing opposite to {\bf B}, which is much higher energy,
the field constraint is
\be
\label{yparb2}
\mu B < -2(3\zeta_R-K_1), \quad p=-1.
\ee
One can also consider deviations at $qa=\pi$, which might connect the state to $y$-alternating structure,
but that gives constraints already satisfied by the requirements from $\lambda_{\phi}(0)>0$.
There is no tendency for $y$-parallel to destabilize into an alternating structure.

\begin{figure}
\includegraphics[width=\figwidth,angle=0]{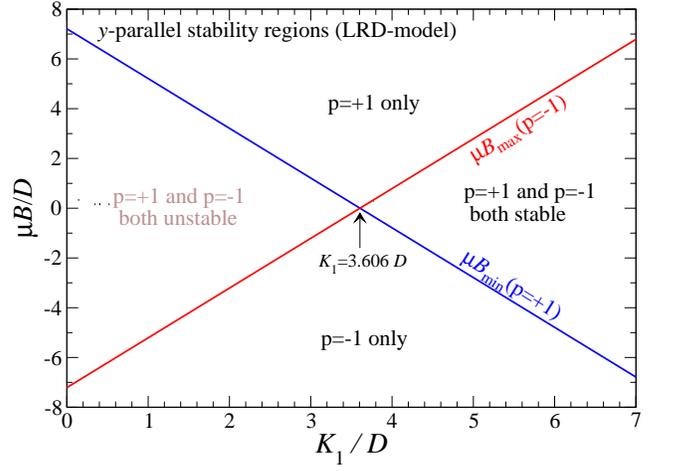}
\caption{\label{muB_lrd_ypar} For a $y$-parallel state with LRD interactions ($R\to \infty$), the field
$\mu B$ required for stabilization, as a function of the in-plane anisotropy $K_1$, taking $K_3=0$. The minimum
field required for $p=+1$ appears in Eq.\ (\ref{yparb1}); the maximum field for $p=-1$ 
appears in Eq.\ (\ref{yparb2}), where $p$ indicates the polarization direction ($\pm \hat{y}$).}
\end{figure}

The resulting stable regions for $y$-parallel states with infinite-range dipole interactions
are depicted in Fig. \ref{muB_lrd_ypar}.  Around a central point at $K_1 \approx 3.606 D$, $B=0$, 
there are four distinct regions: 
a forbidden region with $K_1<3.606 D$ where neither is stable, 
two exclusive regions where only one of the polarizations is stable, 
and a bistable region with $K_1> 3.606 D$ where both polarization are stable.

\subsection{Linearized analysis of $y$-alternating states}
\label{Lin-yalt}
For $y$-alternating states, the analysis requires a two-sublattice model: sites at even (odd) $n$ are 
considered to be on the A (B) sublattice.  When a field is applied, symmetry is broken, making the two 
sublattices inequivalent, and the theory requires different deviation waves on the two sublattices.  
Further, the dynamic frequency eigenvalues now are not given by Eq.\ (\ref{wqll}), but another expression, 
due to the symmetry breaking by the field. 

\subsubsection{Expanding $H$ for $y$-alternating states}
For $y$-alt states, the equilibrium in-plane angles on the two sublattices are $\phi_A^{(0)}=p\frac{\pi}{2}$
and  $\phi_B^{(0)}=-p\frac{\pi}{2}$, where polarization $p=\pm 1$ to give two $y$-alt states.  
With deviations, the in-plane angles are replaced by 
\be
\phi_n \rightarrow (-1)^n p \frac{\pi}{2}+\phi_n,
\ee
where $\phi_n$ are now the small deviations from the equilibrium $y$-alt state. With $\theta_n$
being the out-of-plane deviation at a site, the contributions to $H$ can be expanded to quadratic order
in the angles.  

Consider dipolar interactions. Site $n$ is on one sublattice.  Then site $n+k$ with $k$ odd
belongs to the other sublattice ($n,n+k$ is an AB bond). Their pair contribution to $H$, up to quadratic order, 
is found to be
\begin{align}
H_{n,k}^D & \approx  \frac{D}{k^3} \big(-1+\tfrac{1}{2}\phi_n^2+\tfrac{1}{2}\phi_{n+k}^2 +2 \phi_n \phi_{n+k}
\nonumber \\
& +\tfrac{1}{2}\theta_n^2+\tfrac{1}{2}\theta_{n+k}^2+\theta_n\theta_{n+k}\big), \quad k=\text{odd}.
\end{align}
On the other hand, when $k$ is even, the pair is on the same sublattice (AA or BB bonds), and the contribution is
different,
\begin{align}
H_{n,k}^D & \approx  \frac{D}{k^3} \big(1-\tfrac{1}{2}\phi_n^2-\tfrac{1}{2}\phi_{n+k}^2 -2 \phi_n \phi_{n+k}
\nonumber \\
& -\tfrac{1}{2}\theta_n^2-\tfrac{1}{2}\theta_{n+k}^2+\theta_n\theta_{n+k}\big), \quad k=\text{even}.
\end{align}
In the full Hamiltonian, each $(n,n+k)$ pair should be counted once. 
These are equivalent to a {\em single} expression for any $n,k$, with $k\ge 1$, whose sum contributes to $H$,
\begin{align}
H_{n,k}^D \approx \frac{D}{k^3} \big[(-1)^k\big(1-\phi_n^2 & -2\phi_n\phi_{n+k}-\theta_n^2\big)
\nonumber \\
& +\theta_n\theta_{n+k}\big], \quad k\ge 1.
\end{align}
There are also anisotropy terms,
\be
H_n^K \approx -K_1(1-\phi_n^2-\theta_n^2)+K_3 \theta_n^2.
\ee
The field terms for both sublattices can be expressed succinctly, 
\be
H_n^B \approx -(-1)^n p\mu B (1-\tfrac{1}{2}\phi_n^2-\tfrac{1}{2}\theta_n^2).
\ee
The combination of dipolar, anisotropy and field terms produces the per-site Hamiltonian, which is 
composed from zeroth and second order parts: $H_n \approx H_n^{(0)}+H_n^{(2)}$.
The zeroth order term reproduces the per-site equilibrium energy $u_{y\text{-alt}}$ in (\ref{uyalt}), 
independent of the field,
\be
H_{n}^{(0)} = D\, \sum_{k=1}^R \frac{(-1)^k}{k^3}-K_1 = D{\rm Cl}_{R,3}(\pi)-K_1.
\ee
There remains the quadratic parts,
\begin{align}
H_n^{(2)} = D \sum_{k=1}^R  \frac{1}{k^3} & \big[(-1)^k \left(-\phi_n^2-2\phi_n \phi_{n+k}-\theta_n^2\right)
\nonumber \\
& +\theta_n\theta_{n+k}\big] 
+K_1(\phi_n^2+\theta_n^2)+K_3 \theta_n^2 \nonumber \\
& +(-1)^n \tfrac{1}{2} p\mu B (\phi_n^2+\theta_n^2).
\end{align}
Using the matrix notation of Eq.\ (\ref{qf}), the on-site matrix elements alternate by site due to the
field,
\begin{align}
\label{MM}
M_{\phi,n,n} & = -D {\rm Cl}_{R,3}(\pi)+K_1+(-1)^n\tfrac{1}{2} p\mu B, \nonumber \\
M_{\theta,n,n} & = -D {\rm Cl}_{R,3}(\pi)+K_{13}+(-1)^n\tfrac{1}{2} p\mu B.
\end{align}
The inter-site elements come from the pair terms in $H_n^{(2)}$,
\be
\label{MMi}
M_{\phi,n,n+k} = -D\frac{(-1)^k}{k^3}, \quad M_{\theta,n,n+k} = \frac{D}{2k^3}.
\ee
These resemble the $y$-parallel matrix elements, except for the alternation with separation $k$ in $M_{\phi,n,n+k}$.

\subsubsection{Frequency eigenvalues of $y$-alternating states}
Although the energy eigenvalues of $\bm{M}_{\phi}$ and $\bm{M}_{\theta}$ can be found, they do not play
a direct role in the expressions for the frequency eigenvalues of $y$-alternating states. Instead,
we inspect the dynamics that results from (\ref{LRDmotion}), assuming inversion symmetry and separating out
the on-site interaction,
\begin{align}
-\ii \omega \frac{\mu}{\gamma_{\rm e}} \phi_n &= 
2\big\{ M_{\theta,n,n}\theta_n+\sum_{k=1}^R M_{\theta,n,n+k}(\theta_{n-k}+\theta_{n+k}) \big\}, \nonumber \\
-\ii \omega \frac{\mu}{\gamma_{\rm e}} \theta_n &= 
-2\big\{ M_{\phi,n,n}\phi_n+\sum_{k=1}^R M_{\phi,n,n+k}(\phi_{n-k}+\phi_{n+k}) \big\}. 
\end{align}
A two-sublattice traveling wave expression is assumed,  
\be
(\phi_n,\theta_n) = \begin{cases} (a_{\phi},a_{\theta}) e^{\ii (q n a-\omega t)} & n=\text{even, A-sites}, \\
(b_{\phi},b_{\theta}) e^{\ii (q n a-\omega t)} & n=\text{odd, B-sites}. \end{cases}
\ee
Substituted into (\ref{LRDmotion}),  there results a pair of coupled $2\times 2$ matrix equations, 
similar to those appearing in analysis of remanent states of square-lattice ASI \cite{Wysin23},
\begin{align}
\label{w2x2}
-\ii\omega \left(\begin{array}{c} a_{\phi} \\ b_{\phi} \end{array} \right)
& = \left( \begin{array}{cc} m_{aa} & m_{ab} \\ m_{ba} & m_{bb} \end{array} \right) 
\left(\begin{array}{c} a_{\theta} \\ b_{\theta} \end{array} \right),
\nonumber \\
-\ii\omega \left(\begin{array}{c} a_{\theta} \\ b_{\theta} \end{array} \right)
& = -\left( \begin{array}{cc} n_{aa} & n_{ab} \\ n_{ba} & n_{bb} \end{array} \right) 
\left(\begin{array}{c} a_{\phi} \\ b_{\phi} \end{array} \right).
\end{align}
The elements of the $2\times 2$ matrices $\bm{m}$ and $\bm{n}$ come from projecting $\bm{M}_{\theta}$ and 
$\bm{M}_{\phi}$ onto the two-sublattice traveling wave, such as the matrix $\bm{m}$ due to $\theta$ variations,
\begin{align}
m_{aa} & = 2 \frac{\gamma_{\rm e}}{\mu} \big[ M_{\theta,AA}+\sum_{k=\text{even}}^R 2M_{\theta,n,n+k} \cos kqa\big], 
\nonumber \\
m_{bb} & = 2 \frac{\gamma_{\rm e}}{\mu} \big[ M_{\theta,BB}+\sum_{k=\text{even}}^R 2M_{\theta,n,n+k} \cos kqa\big], 
\nonumber \\
m_{ab} & = m_{ba} = 2 \frac{\gamma_{\rm e}}{\mu}  \sum_{k=\text{odd}}^R 2M_{\theta,n,n+k} \cos kqa.
\end{align}
The symbols $M_{\theta,AA}$ and $M_{\theta,BB}$ indicate the on-site matrix elements for each sublattice (they
differ in the field term).  The sums, due to dipole pair interactions, are over positive values of $k$ and are 
independent of the choice of a central site $n$.  The cosines result from adding interactions in both
directions,
\be
e^{\ii qka}+e^{\ii q(-k)a} = 2\cos kqa.
\ee
Using the known matrix elements of $\bm{M}_{\theta}$ from (\ref{MM}) and (\ref{MMi}), these are
\begin{align}
m_{aa} &= 2\big\{D[{\rm Cl}_{3{\rm e}}(qa)-{\rm Cl}_{3}(\pi)]+K_{13}+\tfrac{1}{2}p\mu B \big\}, \nonumber \\
m_{bb} &= 2\big\{D[{\rm Cl}_{3{\rm e}}(qa)-{\rm Cl}_{3}(\pi)]+K_{13}-\tfrac{1}{2}p\mu B \big\}, \nonumber \\
m_{ab} &= m_{ba} = 2D\, {\rm Cl}_{3{\rm o}}(qa) ,
\end{align}
which depend on even-term and odd-term Clausen sums, 
\be
{\rm Cl}_{3{\rm e}}(qa) = \sum_{k={\rm even}}^{R} \frac{\cos kqa}{k^3},
\quad {\rm Cl}_{3{\rm o}}(qa) = \sum_{k={\rm odd}}^{R} \frac{\cos kqa}{k^3}.
\ee
The matrix $\bm{n}$ due to $\phi$ variations has similar structure, 
\begin{align}
n_{aa} & = 2 \frac{\gamma_{\rm e}}{\mu} \big[ M_{\phi,AA}+\sum_{k=\text{even}}^R 2M_{\phi,n,n+k} \cos kqa\big], 
\nonumber \\
n_{bb} & = 2 \frac{\gamma_{\rm e}}{\mu} \big[ M_{\phi,BB}+\sum_{k=\text{even}}^R 2M_{\phi,n,n+k} \cos kqa\big], 
\nonumber \\
n_{ab} & = n_{ba} = 2 \frac{\gamma_{\rm e}}{\mu}  \sum_{k=\text{odd}}^R 2M_{\phi,n,n+k} \cos kqa.
\end{align}
Using the matrix elements from (\ref{MM}) and (\ref{MMi}), these become
\begin{align}
n_{aa} &= 2\big\{ D[-2{\rm Cl}_{3{\rm e}}(qa)-{\rm Cl}_3(\pi)]+K_1+\tfrac{1}{2}p\mu B \big\}, \nonumber \\
n_{bb} &= 2\big\{ D[-2{\rm Cl}_{3{\rm e}}(qa)-{\rm Cl}_3(\pi)]+K_1-\tfrac{1}{2}p\mu B \big\}, \nonumber \\
n_{ab} &= n_{ba} = 4D\, {\rm Cl}_{3{\rm o}}(qa) .
\end{align}

\begin{figure}
\includegraphics[width=\figwidth,angle=0]{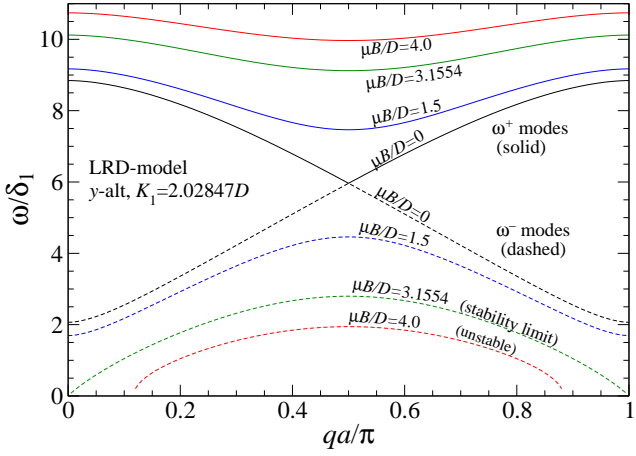}
\caption{\label{walrd-20} For a $y$-alternating state in the LRD-model, typical dispersion relations
for $K_1/D=2.02847$, $K_3=0$, at indicated field strengths $\mu B$, as obtained from Eq.\ \ref{awesome}.
Note the instability when $\mu B$ surpasses its upper allowed value of $3.1554D$ (the triple point for
the three phases, see Fig.\ \ref{phases_lrd}), beyond which the state should transform into $y$-parallel.}
\end{figure}

The frequency eigenvalues can be obtained either by eliminating say, the $(a_{\theta},b_{\theta})$ amplitudes
from (\ref{w2x2}), and solving a $2\times 2$ reduced system, or,  keeping all four amplitudes and solving the
$4\times 4$ eigenvalue problem equivalent to (\ref{w2x2}), expressed as
\be
\left(\begin{array}{cc} \begin{array}{cc} 0 & 0 \\ 0 & 0 \\ \end{array} & \bm{m} \\ \bm{n} &
 \begin{array}{cc} 0 & 0 \\ 0 & 0 \\ \end{array}  \end{array} \right)
\left( \begin{array}{c} a_{\phi} \\ b_{\phi} \\ \ii a_{\theta} \\ \ii b_{\theta} \end{array} \right)
= \omega
\left( \begin{array}{c} a_{\phi} \\ b_{\phi} \\ \ii a_{\theta} \\ \ii b_{\theta} \end{array} \right).
\ee
It is a straightforward exercise \cite{Wysin23} to obtain the eigenvalues of the $4\times 4$ matrix 
on the LHS, which are given from
\be
\label{awesome}
(\omega^{\pm})^2 = \tfrac{1}{2}\left[(\bm{m}^{\dagger}\cdot\bm{n})
\pm\sqrt{(\bm{m}^{\dagger}\cdot\bm{n})^2-4|\bm{m}||\bm{n}|}\right].
\ee
The dots indicate term-by-term scalar products of the $2\times 2$ matrices 
($\bm{m}^{\dagger}\cdot\bm{n}=\sum_{i,j}m_{ij}n_{ji}$), and $|\bm{m}||\bm{n}|$ is the product of their determinants.
We calculate the two frequencies with positive real parts, which give waves traveling in the positive $x$-direction.
When the applied field $B$ is zero, it is  possible to show that the frequency eigenvalues are given by 
an expression equivalent to Eq.\ (\ref{wqll}), namely,
\be
(\omega^{\pm})^2 = \lambda_{\bm{m}}^{\pm}\lambda_{\bm{n}}^{\pm},
\ee
where the eigenvalues of the $\bm{m}$ and $\bm{n}$ matrices enter on the RHS.  Once the field is nonzero,
however, this form does not hold. A simple expression in terms of the energy eigenvalues has not 
been found.   

\begin{figure}
\includegraphics[width=\figwidth,angle=0]{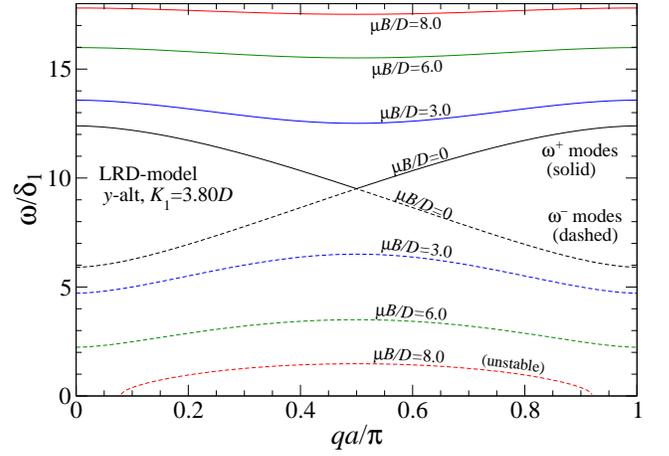}
\caption{\label{walrd-38} For a $y$-alternating state in the LRD-model, typical dispersion relations
for $K_1/D=3.80$, $K_3=0$, at indicated field strengths $\mu B$, as obtained from Eq.\ \ref{awesome}.
Instability occurs for $\mu B>7.73147 D$, see Eq.\ (\ref{muB-lrd-yalt}) and Fig.\ \ref{phases_lrd},
beyond which the state should transform into $y$-parallel.}
\end{figure}

Some typical dispersion relations are shown in Fig.\ \ref{walrd-20}, for the triple point anisotropy value, $K_1/D=2.02847$,
which has instability for $\mu B/D > 3.1554$. The instability is driven both at $qa=0$ and at $qa=\pi$, and 
should transform the state into $y$-parallel once $\mu B$ is above this limit.  A similar behavior appears 
for $K_1/D=3.80$ as shown in Fig.\ \ref{walrd-38}, which allows for much stronger field before instability takes place.  
Note that the frequencies do not depend on the phase-like parameter $p$, which only determines whether even or odd 
sites are aligned with the field.

\subsubsection{General stability of $y$-alternating states}
%
\begin{figure}
\includegraphics[width=\figwidth,angle=0]{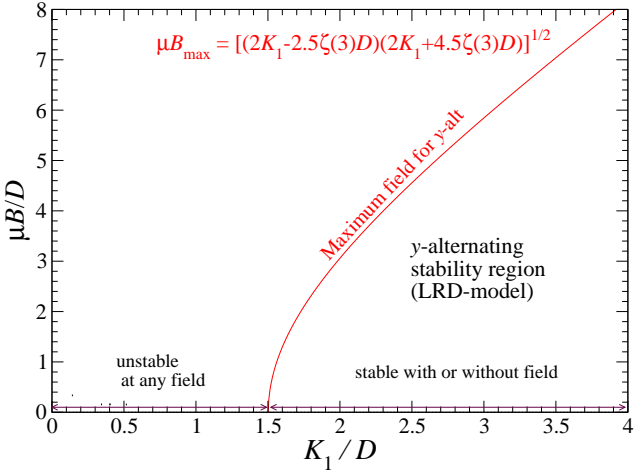}
\caption{\label{muB_lrd_yalt} For a $y$-alternating state with LRD interactions, the maximum applied field
strength $\mu B$ below which the state is stable, and overall stability region,
as a function of relative in-plane anisotropy $K_1/D$, as given in Eq.\ (\ref{muB-lrd-yalt}).
This is based on requiring positive energy eigenvalues, especially $\lambda_{\bm{n}}^{-}>0$.}
\end{figure}
If any frequency eigenvalue from Eq.\ (\ref{awesome}) for any value of $q$ becomes negative or imaginary, 
the $y$-alternating state is unstable.  However, it is mathematically difficult to apply this principle
for determining the range of applied field for which the state will remain stable.

Instead, we again use the principle that real and positive energy eigenvalues are required for stability. Instability
due to deviations in $\phi$ or $\theta$ is indicated by negative energy eigenvalues, signifying that the state 
can lower its energy and destabilize to a different state.  The point where an eigenvalue is zero gives a limiting 
value for the applied field, for the given anisotropy values.  This principle can be applied to the $\bm{m}$ and $\bm{n}$ 
matrices, even though it has not been possible to write the frequencies in terms of energy eigenvalues
for $y$-alt states.

The matrices $\bm{m}$ and $\bm{n}$ have this form which highlights the field dependence:
\be
\bm{n} = \left( \begin{array}{cc} n_{aa} & n_{ab} \\ n_{ba} & n_{bb} \end{array} \right)
=\left( \begin{array}{cc} n_{aa}^0+p\mu B & n_{ab} \\ n_{ba} & n_{aa}^0-p\mu B  \end{array} \right),
\ee
where $=n_{aa}^0$ is the zero-field diagonal matrix element, and $n_{ab}=n_{ba}$ is independent of the field.
A brief calculation gives the two eigenvalues,
\be
\lambda_{\bm{n}}^{\pm} = n_{aa}^0 \pm \sqrt{n_{ab}^2+(\mu B)^2},
\ee
with a similar expression for the pair of $\lambda_{\bm{m}}^{\pm}$ eigenvalues.
Then the stability requirement $\lambda_{\bm{n}}^{\pm}>0$ gives a constraint on the applied field magnitude,
\be
\mu B < \sqrt{\left(n_{aa}^0 \right)^2-n_{ab}^2}.
\ee
Inspecting the typical dispersion relations in the examples in Figs. \ref{walrd-20} and \ref{walrd-38}, the
instability is initiated equally at $q=0$ and at $qa=\pi$.  At $q=0$, the sums needed in the matrix elements 
for infinite range interactions are
\begin{align}
{\rm Cl}_{3{\rm e}}(0) &= \sum_{k=\text{even}}\frac{1}{k^3}=\sum_{n=1}^{\infty}\frac{1}{(2n)^3}=\tfrac{1}{8}\zeta(3),
\nonumber \\
{\rm Cl}_{3{\rm o}}(0) &= \sum_{k=\text{odd}}\frac{1}{k^3}={\rm Cl}_{3}(0)-{\rm Cl}_{3{\rm e}}(0)=\tfrac{7}{8}\zeta(3).
\end{align}
Then the matrix elements for $q=0$ are found to be
\be
n_{aa}^0 = \zeta(3) D+ 2K_1, \qquad n_{ab} = \tfrac{7}{2}\zeta(3) D.
\ee
Then this implies a requirement on the applied field, for the LRD model,
\be
\label{muB-lrd-yalt}
\mu B < 
\sqrt{\left(2K_1-\tfrac{5}{2}\zeta(3)D\right)\left(2K_1+\tfrac{9}{2}\zeta(3)D\right)}.
\ee
The same result is obtained from $\bm{n}(\pi)$.  The calculation can also be repeated for the eigenvalues 
$\lambda_{\bm{m}}^{\pm}$, which leads to a second constraint on the field, 
\be
\mu B < \sqrt{2K_{13}\left(2K_{13}+\tfrac{7}{2}\zeta(3)D\right)}.
\ee
However, the limiting field due to $\lambda_{\bm{n}}^{\pm}$ is smaller and more restrictive than this, 
and it determines stability.  Therefore, the maximum field for stable $y$-alt solutions is given by Eq.\ 
(\ref{muB-lrd-yalt}). That result is plotted in Fig.\ \ref{muB_lrd_yalt}. Also, $y$-alt states require 
a minimum in-plane anisotropy; stability is only possible in the LRD model if 
\be
K_1 > \tfrac{5}{4}\zeta(3) D.
\ee
If the applied field falls somewhere in the region above or to the left of the maximum allowed curve in Fig.\ 
\ref{muB_lrd_yalt}, then the possible stable states it can transform into are either oblique (for lower 
anisotropy $K_1$) or $y$-parallel (for larger values of $K_1$).  This statement is made more precise in the
next section.

\section{Summary on stability and transformations}
%
\begin{figure}
\includegraphics[width=\figwidth,angle=0]{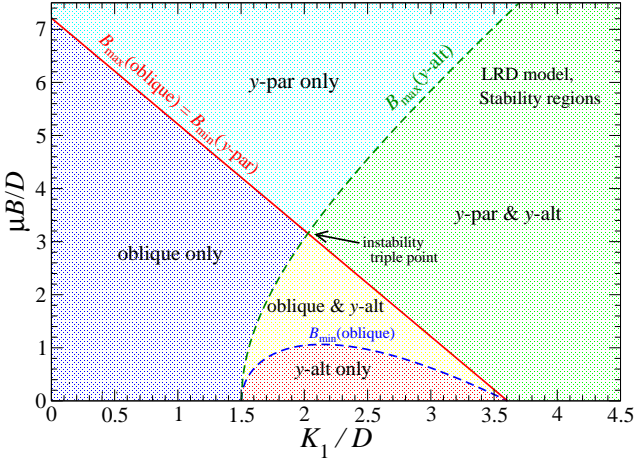}
\caption{\label{phases_lrd} Diagram of the field/anisotropy regions of the LRD-model where the different states are
stable, as determined from having positive energy eigenvalues, especially $\lambda_{\bm{n}}^{-}$. Solid
curves indicate an imperative change of state across the curve (oblique/$y$-par); dashed curves indicate
that only one state is unstable when crossing that curve.}
\end{figure}
Stability has been determined for the three states studied (oblique, $y$-par and $y$-alt) based on requirements
of positive energy eigenvalues, for in-plane and out-of-plane dipole fluctuations.  The stable ranges of 
applied field combined with anisotropy $K_1$ have been determined in the LRD model ($R\to \infty$), for the 
particular case of vanishing easy-plane anisotropy, $K_3=0$.  The stable regions are indicated in a single 
phase-like diagram in Fig.\ \ref{phases_lrd} in the field-anisotropy ($\mu B$-$K_1$) plane.
Each state has an exclusionary region where it is the only stable state: oblique state only for low $B$ and low $K_1$, 
$y$-par only for high $B$ and low $K_1$, and $y$-alt only for low $B$ and intermediate $K_1$.
The states also have metastable regions shared with another of the states.  There are only dual-state regions 
and no triple-state region.  Notably, oblique and $y$-par are mutually exclusive, except along their boundary 
(the solid red line). 

Crossing any of the curves in Fig.\ \ref{phases_lrd} indicates a possible transformation of the system, due to 
instability, from one of the states to another.  Dashed curves are used to indicate that only one of the states 
becomes unstable crossing the curve, while the other one is stable on both sides.  If the state is allowed on 
both sides of a curve, then it does not transform when crossing the curve. 

Crossing over the solid red line in Fig.\ \ref{phases_lrd}, oblique must transform into $y$-par or \textit{vice-versa} 
in the other direction. This conclusion is reached because these processes 
destabilize the original state with fluctuations at $q=0$, see the dispersion relations in Figs.\ \ref{wxlrd-10}
and \ref{wylrd-10}.  Oblique will not transform to $y$-alt by crossing over the solid red line, because it does
not have a destabilizing fluctuation there at $qa=\pi$.  Instead, transformation from oblique to $y$-alt \textit{can} 
take place by crossing downward over the blue dashed curve (minimum field for oblique state stability), 
which occurs with a $qa=\pi$ fluctuation, as seen in Fig.\ \ref{wxlrd-20}.  One could also begin in a $y$-alt 
state and cross over the green dashed curve, transforming either into oblique or $y$-par.  

There is a central point where the red line crosses the green curve in Fig.\ \ref{phases_lrd} where all the phases
become unstable as the point is crossed, which might be termed an \textit{instability triple point}.  Starting in oblique 
and moving through that point gives $y$-par. Starting in $y$-par and moving through the point gives back oblique. Starting
in $y$-alt and moving through the point vertically gives $y$-par, and moving through horizontally gives oblique.
The point is located by setting the expressions for the two curves equal, which gives
\be
\frac{K_1}{D} = \tfrac{189}{112}\zeta(3) \approx 2.02847, \quad
\frac{\mu B}{D} = \tfrac{147}{56} \zeta(3) \approx 3.1554.
\ee
It should be a point of strong fluctuations, as the system cannot easily decide which state to choose.

\begin{figure}
\includegraphics[width=\figwidth,angle=0]{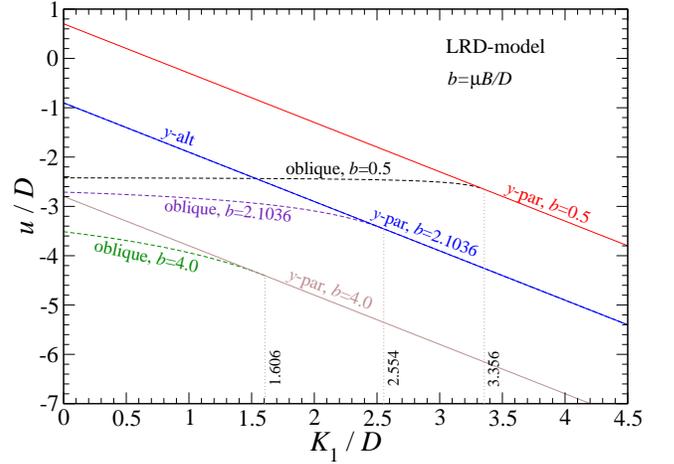}
\caption{\label{u7-lrd} Per-site energies of the states with LRD, from Eqs.\ (\ref{uxp}), (\ref{uyp}), and (\ref{uya}),
for indicated dimensionless field values $b=\mu B/D$, versus in-plane anisotropy, with $K_3=0$. Note that oblique becomes 
the same as $y$-parallel where the two curves meet.   The $y$-alt energy does not depend on $b$. The 
$y$-alt line matches the $y$-par energy line for $b=2.1036$, where those two states are equal in energy for all $K_1$.}
\end{figure}
%
\section{Comparing the states' energies per site}
Due to metastability in this system, comparison of energy per site between states is not a good 
indicator of stability.  An idea of this is given in Fig.\ \ref{u7-lrd} for the LRD model, which shows the states' 
energies per site $u$ vs. anisotropy $K_1$, for different field values, similar to Fig.\ 9 for $B=0$ in Ref.\ 
\cite{Wysin22}. 
These are horizontal scans in the $(K_1,\mu B)$ phase diagram, Fig.\ \ref{phases_lrd}, distinguished by dimensionless 
field $b\equiv \mu B/D$.  The $y$-alt energy does not depend on the field, hence it has only one (straight-line) curve.  
The oblique state energy curves meet with the corresponding $y$-par curves at the maximum $K_1$ where there is a mandatory 
transformation into $y$-par.  There is nothing in the curves to indicate where $y$-alt is stable or unstable, even 
though we know it has a limited stability range.  The $y$-alt line matches the $y$-par line for 
$\mu B =(7/4)\zeta D \approx 2.1036 D$.

\begin{figure}
\includegraphics[width=\figwidth,angle=0]{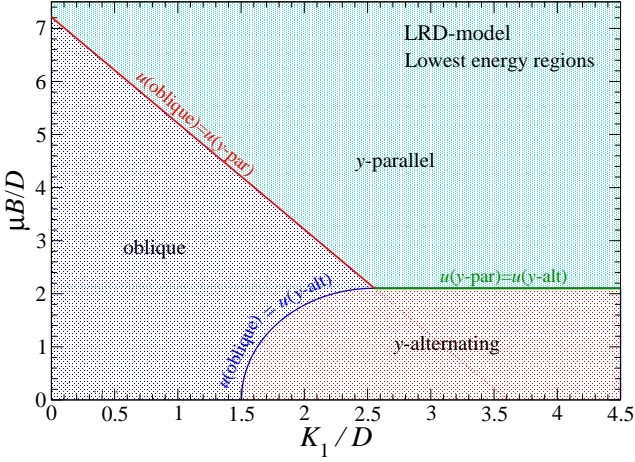}
\caption{\label{lowest_lrd} Diagram of the field/anisotropy regions of the LRD model where the
different states have the lowest energy per site $u$, from Eqs.\ (\ref{uxp}), (\ref{uyp}), and (\ref{uya}).
Solid curves indicate where the primary states have equal energies;  dashed curves indicate where
secondary states have equal energies in that region. Oblique does not exist to the right
of the solid and dotted red lines.} 
\end{figure}

In a better comparison, the regions where each of the three states has the lowest energy are determined,
including LRD interactions, in contrast to the phase-like stability diagram in Fig.\ \ref{phases_lrd} based on 
positive energy eigenvalues.  For example, the region where $u_{\text{oblq}} \le u_{y\text{-par}}$ for polarization
$p=1$ is found to be bounded by 
\be
\label{bx2-lrd}
\mu B = 2(3\zeta D-K_1),
\ee
which is the (red line) boundary in Fig.\ \ref{phases_lrd} based on energy eigenvalues.  However, 
the region where
$u_{\text{oblq}} \le u_{y\text{-alt}}$ is found to be
\be
\mu B \ge 2\sqrt{(K_1-\tfrac{5}{4}\zeta D)(3\zeta D-K_1)}, 
\ee
and this is different from the (dashed blue) minimum field needed to stabilize oblique states.
Finally, for $y$-par with $p=1$, the region where $u_{y\text{-par}} \le u_{y\text{-alt}}$ is given simply by
\be
\mu B \ge \tfrac{7}{4}\zeta D \approx 2.1036 D. 
\ee
These results are combined into a diagram in Fig.\ \ref{lowest_lrd} which shows the 
various regions where each state, including LRD interactions, has the lowest energy per site.
In each of the three regions, there are at least two of the states possible.  The only restriction, 
in principle,  is that oblique and $y$-par are mutually exclusive.  This diagram is notably different 
from the stability diagram in Fig.\ \ref{phases_lrd}.  Having the lowest energy alone does not determine stability.  
Stability is determined, indeed, by having finite frequency dynamic fluctuations over the whole range of
allowed wavevectors.

\begin{figure}
\includegraphics[width=\figwidth,angle=0]{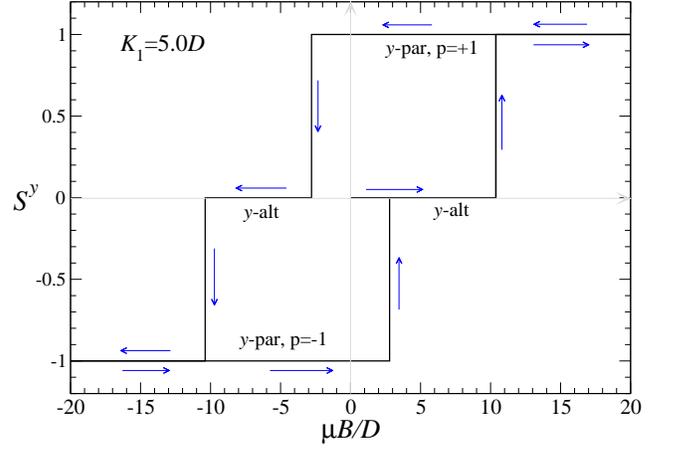}
\caption{\label{my-k5} The dimensionless magnetization per site $S^y$ along the field direction, versus 
dimensionless applied field, for $K_1=5 D$, $K_3=0$, starting from a $y$-alt state at $B=0$, and then cycling 
$B$ through positive and negative values. This is obtained from the LRD model's stability diagram, Fig.\ \ref{phases_lrd},
using the $y$-alt stability limit, Eq.\ (\ref{muB-lrd-yalt}), and the $y$-par stability limits, Eqs.\ (\ref{yparb1})
and (\ref{yparb2}).
}
\end{figure}
\section{Effects on magnetization}
The stability diagram in Fig.\ \ref{phases_lrd} can be used to estimate how the dimensionless magnetization ($S^y$) 
behaves with applied field, and how that response depends on the anisotropy $K_1$. 
Measurement of the magnetization curves versus $B$ could indicate the transitions among the states. 
The entire chain is assumed to be in one of the uniform states (after any transients).
$K_3=0$ is assumed although any positive value will produce similar results.

For $K_1 < 1.50257 D$ and $B=0$, initially $S^y=0$. As $B$ increases positively, $S^y$ will increase linearly
as in Eq.\ (\ref{Sny}), and the system will transform only between an oblique state and a $y$-par state.  
The transition occurs when the dipoles in the oblique state have rotated to be exactly parallel with {\bf B}, 
and $S^y \rightarrow 1$.  There will be no hysteresis when $B$ is reduced back through zero; the transitions 
are reversible.

For $1.50257 < K_1 < 3.606$, the system will naturally be in $y$-alt at $B=0$, which is the lowest energy,
with $S^y=0$.  As $B$ is increased, eventually $y$-alt becomes unstable (at the green dashed curve in Fig.\ 
\ref{phases_lrd}, Eq.\ \ref{muB-lrd-yalt}) and there will be a transition either into an oblique state or $y$-par. 
That depends on where $K_1$ falls relative to the instability triple point. There will be a discontinuous change in
$S^y$. The system will exhibit hysteresis when the field is later dropped back to zero and negative values.   

For $K_1 > 3.606 D$, which applies to more elongated islands, an example magnetization curve is shown 
in Fig.\ \ref{my-k5}, for $K_1=5.0 D$, $K_3=0$.  The system might start in the lowest energy state, $y$-alt, at $B=0$, 
with $S^y=0$. It would stay in that state until $B$ reaches the upper limit for $y$-alt stability 
[Eq.\ (\ref{muB-lrd-yalt})], whereupon it transforms into $y$-par, with $S^y=1$.       
Upon subsequent reduction of $B$ to zero and negative values, there will be other discrete
jumps and hysteresis.   Note how the chain essentially becomes a three-level system in this case.

\section{Conclusions}
This model for a chain of interacting dipoles exposed to a transverse applied field possesses three uniform
states (oblique, $y$-par and $y$-alt) that have metastable properties, depending of the interaction parameters
and the field.
The energy and frequency eigenvalues have been used to determine their stability regions in the
parameter space.
It is to be noted that stability is {\em not} associated with a state being of the lowest energy.
The dynamic modes found give a direct indication of the type of destabilizing transformations 
that would take place when a stability limit is reached or surpassed.

In this problem, dipolar interactions help to stabilize the $y$-alternating states, even where they
coexist with oblique and $y$-par states.
In other systems, dipolar interactions generally influence the magnetic relaxation of chains of magnetic nanoparticles
with randomly oriented anisotropy axes \cite{Anand21}, and induce ordering in two-dimensional lattices
of nanoparticles \cite{Bahiana+04}.
The results here will help to design new dipolar systems with desired switching transitions between metastable states.
Variations on this problem may be especially useful for analysis of states in related systems, such as 2D artificial
spin ices and other interacting dipole structures.

\end{document}